\input{aipcheck}

\documentclass[
    ,final            
  ]
  {aipproc}

\layoutstyle{6x9}

\begin{document}

\title{Star-Planet Interactions}


\author{Evgenya Shkolnik (DTM/CIW), Suzanne Aigrain (Exeter), Steven Cranmer (CfA), Rim Fares (LATT), Malcolm Fridlund (ESA), Frederic Pont (Exeter), J\"urgen Schmitt (Hamburger Sternwarte), Alexis Smith (St.~Andrews), Takeru Suzuki (U.~of Tokyo)}{
  address={}
}



\begin{abstract}

Much effort has been invested in recent years, both observationally and theoretically, to understand the interacting processes taking place in planetary systems consisting of a hot Jupiter orbiting its star within 10 stellar radii. Several independent studies have converged on the same scenario: that a short-period planet can induce activity on the photosphere and upper atmosphere of its host star. The growing body of evidence for such \emph{magnetic} star-planet interactions includes a diverse array of photometric, spectroscopic and spectropolarimetric studies. The nature of which is modeled to be strongly affected by both the stellar and planetary magnetic fields, possibly influencing the magnetic activity of both bodies, as well as affecting irradiation and non-thermal and dynamical processes. \emph{Tidal} interactions are responsible for the circularization of the planet orbit, for the synchronization of the planet rotation with the orbital period, and may also synchronize the outer convective envelope of the star with the planet. Studying such star-planet interactions (SPI) aids our understanding of the formation, migration and evolution of hot Jupiters.

In this proceeding, we briefly summarise the observations and theories presented during the Cool Stars 15 splinter session\footnote{The complete programme with abstracts for the session can be found at
   http://ifa.hawaii.edu/$\sim$shkolnik/SPI\_splinter/Program.html.} of this diverse and growing field of star-planet interactions.
\end{abstract}

\maketitle


\section{Magnetic SPI}

\subsection{SPI as a probe of planetary magnetic fields}

\begin{figure}[h]
\includegraphics[angle=0,width=8cm]{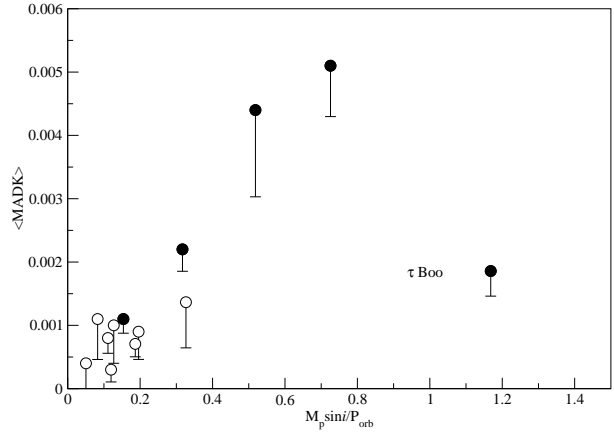}
\caption{The ratio of the minimum planetary mass (in Jupiter masses) to the orbital period (in days) plotted against the average mean-absolute-deviation ($<$MAD$>$) of the Ca II K-line per observing run for 13 stars with close-in planets. The x-axis quantifies the planet's magnetic moment assuming tidal locking, such that $P_{rot}=P_{orb}$. The filled-in circles are stars which exhibit significant night-to-night variability in the Ca II K line: HD 73526, $\upsilon$ And, HD 179949 and HD 189733. The tidally-locked system of $\tau$ Boo does not follow the correlation traced out by the others. This is consistent with a proposed Alfv\'en wave model where the near zero relative motion due to the tidal locking of both the star and the planet produces minimal SPI because of the weak Alfv\'en waves generated as the planet passes through the stellar magnetosphere, thereby transporting little excess energy to the stellar surface along the magnetic field lines \cite{gu05}.  The error bars are one-sided due to the positive contribution of integrated MAD immediately outside the Ca II emission core and reflect the S/N obtained for each target. 
\citep{shko08}}
\label{madk}
\end{figure}

Star-planet interactions (SPI) offer an indirect way to detect and measure planetary magnetic fields, providing the unique opportunity to explore the interiors of planets and set constraints on the rapid hydrodynamic escape of hot Jupiter atmospheres. Observations over the past 6 years reveal evidence suggesting an observable magnetic interaction between a star and its hot Jupiter, which appears as a cyclic variation of stellar activity synchronized to the planet's orbit, e.g.~\cite{shko05,shko08,walk08}. A sample of 13 targets exhibits a tentative correlation between this activity and the ratio of the planet's minimum mass to its rotation period, a quantity proportional to the hot Jupiter's magnetic moment. This intriguing correlation between the planet's magnetic moment and the night-to-night chromospheric activity on its star improves SPI's potential as a probe of extrasolar planetary magnetic fields (Figure~\ref{madk}). 

Direct detections of giant planet magnetic fields, however, require the measurement of radio emission from the planet itself.  Much of the work has so far focused on the massive planet orbiting $\tau$ Boo \cite{bast00,farr03,lazi07,geor07} which placed upper limits comparable to the flux densities predicted from the strongest emissions.  A.~Smith presented the first eclipse observations, which produced upper limits consistent with previous observations.
Based on predictions from a new theoretical model of the cyclotron-maser emission from extrasolar planets \cite{jard08}, observations were carried out of the K-dwarf HD 189733 (with a transiting 1.15-M$_{J}$ planet) \cite{bouc05} with the Green Bank Telescope (GBT) over about 5.5 hours before, during and after secondary eclipse over a bandpass spanning 307--347 MHz. No dip in the radio flux density at the known time of secondary eclipse was observed resulting in a 3-$\sigma$ upper limit to the flux density of 60 mJy. The lack of detection can be explained in three ways:
(1) The frequency of the emission is outside the range of the observations and is determined by the magnetic field strength of the planet.
(2) Geometric effects; the emission could be beamed out of the line-of-sight.
(3) The emission is insufficiently strong to be detected; the strength of the emission is dependent upon the stellar coronal density, which determines how many electrons are available for acceleration.
Hope lies now with the radio wavelength facilities currently under construction, such as the
Square Kilometer Array (SKA) or the Low-Frequency 
Array (LOFAR), which will offer more than one order of magnitude improvement.




\subsection{Variable magnetic SPI}

Several years of spectroscopic observations have shown that both HD 179949 ($P_{orb}$=3.1 d) and $\upsilon$ And ($P_{orb}$=4.6 d) display synchronicity of the Ca II emission with the planet's orbit during several observational epochs while data collected during other seasons exhibit rotational modulation \cite{shko08}. This on/off nature of magnetic SPI in the two systems is likely a function of the changing stellar magnetic field structure throughout its activity cycle. 

\begin{figure}[h]
\includegraphics[angle=0,width=15cm]{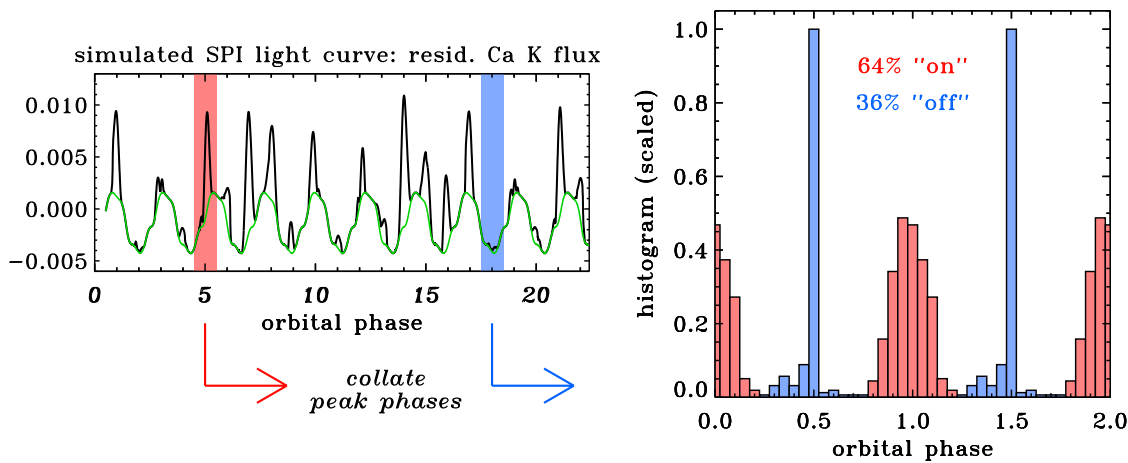}
\caption{Example of simulated Ca II K light curve (left) and ``on/off'' histogram (right) for the parameters of HD 179949.  See text for more details.}
\label{cranmer}
\end{figure}

This is supported by simulations presented by \cite{cran07} of Ca~II K light curves for a close-in extrasolar planet
traversing a series of solar-type multi-pole magnetic fields, which exhibit both
substantial orbit-to-orbit variability and longer-term solar-cycle dependence.  S. Cranmer presented a new statistical analysis of these models at the session.  Figure~\ref{cranmer} illustrates a sample light curve
for the parameters of HD 179949, which shows variations both due to
stellar rotation (green) and SPI field-line enhancements (black).
For a set of 50 consecutive orbits and 11 example years out of a
solar cycle, the phases of peak light were computed for each of the
550 orbits.  About 2/3 of the time, the system was in an ``on'' phase
(red) with peak phases centered on orbital phase\footnote{Orbital phase of 0 is defined by the sub-planetary point.} $\phi_{orb}$=0 and a standard deviation of about
$\pm 0.1$.  When the peak phase was found to be $\pm 0.5$, it was
evident that there was no visible SPI enhancement, and thus the system
was in an ``off'' phase about 1/3 of the time (blue).  The simpler
dipole-like field corresponding to solar minimum gave rise to a
smaller ``on'' fraction (58\%) than at solar maximum (71\%) because
of larger field-line excursions from the point radially below the
planet.  These fractions are similar to the frequency of on/off
phases seen by \cite{shko08}.

\begin{figure}[h]
\includegraphics[angle=0,width=8cm]{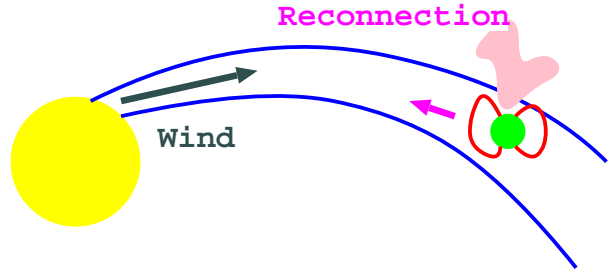}
\caption{Cartoon depicting the magnetic reconnection scenario discussion above.}
\label{suzuki}
\end{figure}

This strong dependence on the star may help, at least partially, explain the rather enigmatic case of the solar-type star orbited by CoRoT Exo-1b at only 5 stellar radii from its host ($P_{orb}=1.5 d$; \cite{barg08}). Spectroscopic followup observations with both HARPS and UVES of this star have demonstrated peculiar flaring properties with strong Na~I, relatively faint (but clear) H$\alpha$, and \emph{no} indication of Ca~II H \& K emission. Spectra taken two days apart exhibit strong yet variable emission strengths of both Na I reversals. Observations 6 weeks later, however, show very faint Na I and no H$\alpha$ emission. Several emission sources have been ruled out: 
(1) an extended background source discounted on the basis of the UVES spectra,
(2) an evaporating planet atmosphere (no sign of variation with period 1.5d and amplitude 370 km/s),
(3) an unknown stellar companion, as it would have to be closer than 500 AU and there is no drift in the velocity curve, and (4) intrinsic stellar activity/flares as no Ca II emission is observed.  The inconclusive results leave open the possibility that reconnection events \emph{between} the star and planet (see Figure~\ref{suzuki}) may produce the Na I emission . Naturally, M. Fridlund et al. plan follow-up observations of this system.


\subsection{Stellar mapping with spectropolarimetry}

It is becoming clearer that stellar fields play an important role in quantitatively studying magnetic SPI. With spectropolarimetry, reconstruction of the large-scale magnetic topologies of hot-Jupiter hosting stars is possible using Zeeman Doppler Imaging (ZDI). ZDI produces tomographic images from a series of circular polarization profiles collected over complete rotational cycles with maximum-entropy image reconstruction to ensure a unique solution (e.g.~Figure~\ref{fares}). It consists of decomposing the field into poloidal and toroidal components, expressed in terms of spherical harmonics expansion. 

In addition to generating a complete magnetic map of a star, the stellar rotation (period and differential rotation) is fully characterized by ZDI. The potential of spectropolarimetric observations in exploring star-planet interactions thus appears extremely promising for comparative observational studies of hot Jupiter systems spanning a range of properties. This will bring constraints of unprecedented quality and completeness to the star-planet interactions models.


The first observational results on $\tau$ Boo, HD 179949 and HD 189733 obtained using ZDI at CFHT/ESPaDOnS and TBL/NARVAL are presented in \cite{cata07,mout07,dona08}, and R.~Fares et al.~(this proceedings). 
$\tau$ Boo shows indications for tidal interactions (discussed in more detail below), whereas HD 179949 may be showing additional evidence of magnetic interactions by varying magnetic field structure as well as by the spectroscopic activity tracers. Determining whether the HD 189733 system is undergoing SPI with this technique has been challenging and requires further observations and analysis in part due to the high level of intrinsic stellar activity on the late-K dwarf. These three are only the beginning as additional spectropolarimetric observations are underway for a variety of hot-Jupiter systems. 


\subsection{Two proposed mechanisms for magnetic SPI}

\paragraph{Magnetic Reconnection Events}

Gu \& Suzuki (in prep.) investigate the effect of magnetic reconnections in the magnetosphere of 
hot Jupiters on the chromosphere and wind of central stars using 1-D MHD simulations.
A fraction of reconnection jets propagate to a central star along stellar 
magnetic field lines. The jets, or the beam of charged particles, interact with 
stellar wind and corona where the density becomes high for sufficient 
Coulomb coupling. 

If the magnetic field strength and the input energy is sufficiently high ($>$0.02G at 10 R$_{\odot}$), warm regions with
temperatures of 10$^4$ -- 10$^6$ K are formed in the chromosphere. They become sources of chromospheric, transition
regional, and coronal radiation fluxes. The radiation fluxes from chromospheric, transition region, and coronal
components are enhanced $\sim$100 times, $\sim$1000 times, and $\sim$10 times, respectively. Moreover, the radiation
fluxes are highly time-dependent because of thermal instability. The calculated chromospheric luminosity is lower than
the observed level (at least for the Ca II emission of HD 179949 \cite{shko05}), possibly because the area at the
foot-point connected to planet magnetosphere is too small. In conjunction with new data indicating SPI and a more
complete accounting of the energy-budgets, these models may explain at least one of the mechanisms behind the phenomenon.

\paragraph{The Jupiter-Io Analogy}

Since the 1960s surface phenomena on Jupiter have been connected with
its moon Io, culminating with the UV detections of Io's (and the other
Galilean moons') footprints on Jupiter.  In situ
measurements of particles and magnetic fields near Io with the Voyager and
Galileo spacecraft have led to a reasonably detailed understanding
of the important physical processes.  The most relevant elements of these are: (1) The strong magnetic field of Jupiter, leading to co-rotation of Jupiter's magnetosphere far beyond Io. (2) The rapid rotation of Jupiter, leading to super-Keplerian magnetospheric rotation at Io's orbit. (3) The mass loading due to Io's volcanic activity, leading to the creation of the Io plasma torus and ensuing low Alfv\'en speeds.  
These properties make Io a MHD generator of currents in excess of 1 million amp producing the observed emissions.  

In the case of extrasolar planets around young stars, conditions (1)-(3) are met and consequently similar processes are expected to occur on far larger scales.  Furthermore, MHD forces might provide a torque to halt planet migration.

Extending the Jupiter-Io mechanism to a hot-Jupiter planetary system, one can estimate the maximally released power through well-known MHD equations, e.g.~\cite{zark01} $\rightarrow$ Up to 10$^{29}$ ergs/s can be released in an Alfv\'en wing around the planet, especially for a close-in giant planet around a rapidly rotating, therefore young, star.  This is further detailed by J.~Schmitt (this proceedings).

\section{Tidal SPI}

\begin{figure}[h]
\includegraphics[angle=270,width=15cm]{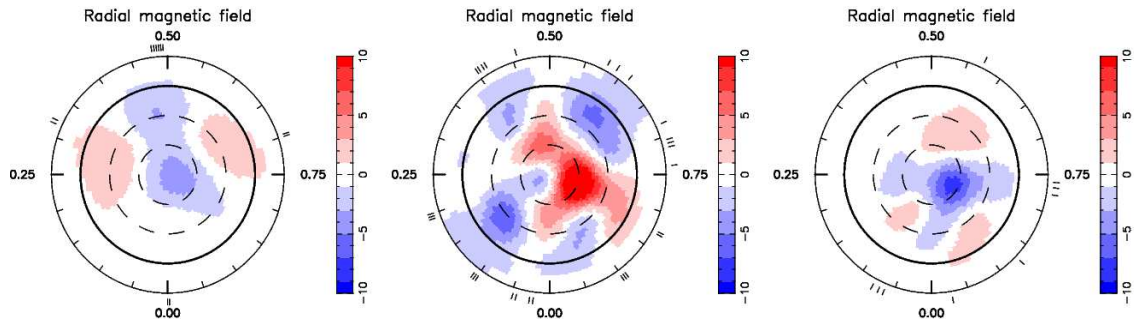}
\caption{Maximum-entropy reconstructions of the large-scale magnetic topology of $\tau$ Boo shows two consecutive polarity switches in 2006 (left), 2007 (middle) and 2008 (right, preliminary map), here illustrated on of the radial component of $\tau$ Boo (\cite{cata07,dona08}, Fares et al.~in prep.) The star is shown in flattened polar projection down to latitudes of --30$^{\circ}$, with the equator depicted as a bold circle and parallels as dashed circles. The same flux scale (labeled in Gauss) is used, and orbital phases ($P_{orb}$=3.3 d) of observations are marked with radial ticks around the plots.}
\label{fares}
\end{figure}

Some hot Jupiters may be close enough to their host star and massive enough to tidally spin up the rotation of the star, potentially forcing the stars outer layers into synchronous rotation with the planet.  This may also strongly impact the star's long-term activity cycle (analogous to the 22-year solar cycle). Evidence for these effects is presented by \cite{cata07} and \cite{dona08}, where $\tau$ Boo is not only shown to be tidally locked to its planet ($P_{orb} = P_{rot}$ = 3.3 d at intermediate latitudes), but also a double polarity reversal was detected in the large-scale stellar magnetic field. Figure~\ref{fares} shows the ZDI reconstructed magnetic map of $\tau$ Boo as observed in 2006, 2007 and 2008, where a polarity switch from year to year is evident (R. Fares). 

\begin{figure}[h]
\includegraphics[angle=0,width=10cm]{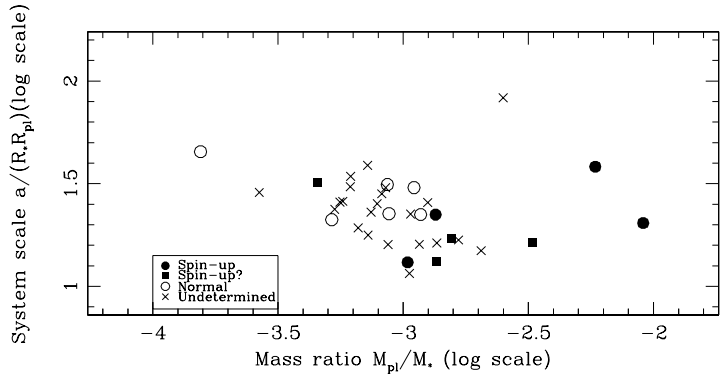}
\caption{The orbital distance in units of stellar radius, and the planet-to-star mass ratio for transiting exoplanets. The strength of tidal effects increases towards the top-left to bottom-right. Filled symbols show host stars which are measured to rotate faster than typical for their spectral type. Their position on this plot may be an indication of tidal spin-up.}
\label{pont}
\end{figure}

Through angular momentum exchange, tidal SPI would also lead to a decay of the orbit of the planet, causing it to spiral towards the star. The known sample of transiting exoplanets is finally large enough to begin to probe this effect. Indeed, there are tentative indications in Figure~\ref{pont} that stars with close-in \emph{and} heavy planets rotate faster than the average stellar rotation speed (F.~Pont). If confirmed, this stellar spin-up and planetary tidal decay may explain the observed correlation between mass and period for hot Jupiters, such that the more massive planets have undergone significant orbital decay on their way towards the star.

\subsection{Tidal interactions as observed by CoRoT}

The CoRoT satellite\footnote{CoRoT was launched in December 2006 and is operated by the French space agency CNES with contributions from Austria, Belgium, Brazil, Germany, Spain and the European Space Agency.} is the first space mission designed to search for planetary transits and will aid in the understanding of both magnetic and tidal SPI. The high precision, high time sampling, long baseline and almost continuous light curves it provides  enable us not only to measure precise planet parameters, but also to measure rotation periods for the host stars, and to study active regions on their surface. In addition to the variable (and mysterious) Na I emission of CoRoT Exo-1b discussed above, several examples have already emerged:

(1) CoRoT-Exo-2 \cite{alon08} is a very active G-star hosting a $3.3\,M_{\rm Jup}$ planet in a 1.75\,d orbit. A discrete auto-correlation function (ACF, Collier-Cameron et al., submitted) of the out-of transit (OOT) light curve measured a rotation period of 4.55$\pm$0.1\,d and showed evidence for two evolving active regions on opposite hemispheres with no detectable differential rotation (S. Aigrain). This may be an indication of induced stellar activity in the tidal bulges on the star generated by the presence of the planet \cite{cunt00}. \cite{lanz08b} find tentative evidence that the star's spotted areas vary at a multiple of the star-planet synoptic period, a possible indication that star-planet \emph{magnetic} interaction is also at play in this system.

(2) With a mass of $21.6\,M_{\rm Jup}$ and an orbital period of 4.26\,d, CoRoT-Exo-3b (Deleuil et al. submitted) lies at the boundary of the planet and brown dwarf regimes, and is a prime candidate for tidal synchronization of its mid-F type host star, a hypothesis supported by the spectroscopically determined rotational velocity of the latter. Unfortunately, the star's activity level is not sufficient to detect a clear photometric rotation period from the light curve. 

\begin{figure}[h]
\includegraphics[angle=0,width=10cm]{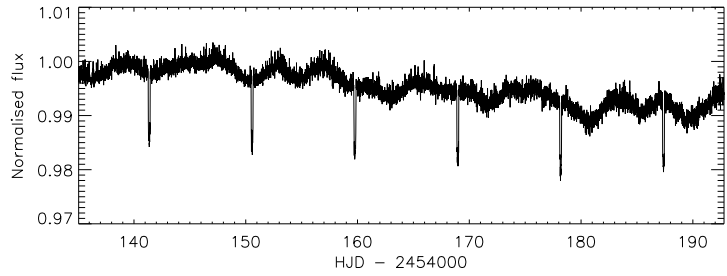}
\caption{Normalised light curve of CoRoT-Exo-4, showing 6 planetary transits and rotational modulation of several emerging and decaying active regions.}
\label{aigrain}
\end{figure}

(3) On the other hand, tidal effects should be much weaker in the CoRoT-Exo-4 system, consisting of a late-F star and a Jupiter-mass planet with a period of 9.21\,d \cite{aigr08,mout08}. Nevertheless, ACF analysis of the OOT light curve (Figure~\ref{aigrain}) indicates that stellar rotation is synchronised with the orbit ($P_{\rm rot}=8.8\pm1.1$\,d), a surprising result which remains to be explained.

\begin{theacknowledgments}

We would like to thank the Cool Stars 15 LOC and SOC for providing an opportunity for this splinter session and for organizing a terrific meeting in St. Andrews.
 
\end{theacknowledgments}



\bibliographystyle{aipproc}   

\bibliography{SPI_refs}

\IfFileExists{\jobname.bbl}{}
 {\typeout{}
  \typeout{******************************************}
  \typeout{** Please run "bibtex \jobname" to optain}
  \typeout{** the bibliography and then re-run LaTeX}
  \typeout{** twice to fix the references!}
  \typeout{******************************************}
  \typeout{}
 }

\end{document}